\documentstyle[11pt,newpasp,twoside,epsf]{article}

\begin{document}

\title{Using Balmer line profiles to investigate convection in A and F stars}

\author{Barry Smalley}
\affil{Astrophysics Group, School of Chemistry \& Physics, Keele University,
Staffordshire, ST5 5BG, United Kingdom}
\author{Friedrich Kupka}
\affil{Institute for Astronomy, University of Vienna,  T\"{u}rkenschanzstr. 17,
A-1180, Austria}
\affil{Astronomy Unit, School of Mathematical Sciences, Queen Mary,
University of London, Mile End Road, London, E1 4NS, United Kingdom}

\begin{abstract}
Balmer lines are an important diagnostic of stellar atmospheric
structure, since they are formed at a wide range of depths within the
atmosphere. The different Balmer lines are formed at slightly different
depths making them useful atmospheric diagnostics. The low sensitivity to
surface gravity for stars cooler than $\sim$8000~K makes them excellent
diagnostics in the treatment of atmospheric convection. For hotter stars
Balmer profiles are sensitive to both effective temperature and surface
gravity. Provided we know the surface gravity of these stars from some other
method (e.g. from eclipsing binary systems), we can use them to determine
effective temperature.

In previous work, we have found no significant systematic problems with using
$uvby$ photometry to determine atmospheric parameters of fundamental (and
standard) stars. In fact, $uvby$ was found to be very good for obtaining both
$T_{\rm eff}$ and $\log g$. Using H$\alpha$ and H$\beta$ profiles, we have
found that both the Canuto \& Mazzitteli and standard Kurucz mixing-length
theory without approximate overshooting are both in agreement to within the
uncertainties of the fundamental stars. Overshooting models were always
clearly discrepant. Some evidence was found for significant disagreement
between {\em all\/} treatments of convection and fundamental values around
8000$\sim$9000~K, but these results were for fundamental stars {\em
without\/} fundamental surface gravities. We have used stars with fundamental
values of both $T_{\rm eff}$ and $\log g$ to explore this region in more
detail.
\end{abstract}



\section{Introduction}

Balmer lines are an important diagnostic of stellar atmospheric structure
since they are formed at a wide range of depths within the atmosphere. The
depth of formation of H$\alpha$ is higher than that of H$\beta$, thus
observations of these profiles provide useful diagnostics (e.g. Gardiner 2000).
Balmer profiles are relatively insensitive to surface gravity for stars
cooler than $\sim$8000~K (Gray 1992, see also Heiter et al. 2002), whilst
sensitive to the treatment of atmospheric convection (e.g. van't Veer \&
M\'{e}gessier 1996, Castelli et al. 1997, Gardiner 2000, Heiter et al. 2002). 
For stars hotter than $\sim$8000~K the profiles are sensitive to both effective
temperature and surface gravity. However, provided we know surface gravity
from some other means (e.g. from eclipsing binary systems), we can use them to
determine effective temperature.

In previous work, Smalley \& Kupka (1997) found no significant systematic
problems with $uvby$ and fundamental (and standard) stars. In fact, $uvby$
was found to be very good for obtaining $T_{\mathrm eff}$ and $\log g$. Using H$\alpha$ and
H$\beta$ profiles, Gardiner et al. (1999) found that both the Canuto \&
Mazzitteli (1991, 1992) and standard Kurucz (1993) mixing-length theory
without overshooting (see Castelli et al. 1997) are both in agreement to
within the uncertainties of the fundamental stars. Overshooting models were
always clearly discrepant. However, Gardiner et al. (1999) found some
evidence for significant disagreement between {\em all\/} treatments of
convection and fundamental values around 8000$\sim$9000~K. In this region the
effects of $\log g$ cannot be ignored, and the majority of the $T_{\mathrm eff}$ stars did
not have fundamental values of $\log g$. We have used binary systems with
fundamental values of $\log g$, determined fundamental values of $T_{\mathrm eff}$ and
compared the results with those from fitting models to Balmer-line profiles.

\section{Effective temperatures of binary systems}

Eclipsing binary systems provide ideal test stars for comparing to models,
since they enable us to obtain fundamental values of $T_{\mathrm eff}$ and $\log g$. We can
obtain fundamental values of $T_{\mathrm eff}$ provided we know the apparent angular
diameter and total integrated (bolometric) flux at the Earth. In the case of
binary systems, where there are no direct measurements of angular diameters,
we can obtain them from the stellar radius and the parallax of the system
(Smalley \& Dworetsky 1995).

Available spectrophotometry was taken from various sources. Unfortunately,
not all the systems have enough high-resolution spectrophotometry. In these
cases, however, we have at least $UBV$ or $UBVRI$ colours, which were used to
estimate the optical fluxes. Near-infrared fluxes were taken from the Gezari
et al. (1987), 2MASS and DENIS catalogues. The lack of available fluxes for
binary systems is something that the Citadel ASTRA Spectrophotometer (Adelman et
al. 2002) could address.

In most cases the $T_{\mathrm eff}$ values obtained using the Hipparcos parallaxes are
in agreement with that obtained from other methods (e.g. Infrared Flux
Method, $uvby$ photometry, etc.) and other determinations (e.g. Jordi et al.
1997, Ribas et al. 1998). However, three systems were anomalous and these are
discussed in detail in Smalley et al. (2002).

\begin{table}
\caption{Fundamental values of $T_{\mathrm eff}$ for binary stars. [Adapted from Smalley et al. (2002)]}
\label{fund_values}
\begin{tabular}{llllll} \tableline
             &     &                   & Sp.   & $f_{\earth}$               &             \\
Star         &     & $\log g$             & Types & (10$^{-6}$ W m$^2$) & $T_{\mathrm eff}$       \\ \tableline

12 Per       &  A  & 4.16  $\pm$ 0.03  & F8V   & 195.  $\pm$ 12.8 & 6371 $\pm$ 176 \\
             &  B  & 4.24  $\pm$ 0.03  & G2V   & 115.  $\pm$ 5.06 & 6000 $\pm$ 143 \\[+1mm]

CD Tau       &  A  & 4.087 $\pm$ 0.010 & F6V   & 24.3  $\pm$ 3.06 & 6110 $\pm$ 415 \\
             &  B  & 4.174 $\pm$ 0.012 & F6V   & 20.7  $\pm$ 1.61 & 6260 $\pm$ 397 \\[+1mm]

UX Men       &  A  & 4.272 $\pm$ 0.009 & F8V   & 7.48  $\pm$ 1.12 & 6171 $\pm$ 302 \\
             &  B  & 4.306 $\pm$ 0.009 & F8V   & 6.52  $\pm$ 0.54 & 6130 $\pm$ 233 \\[+1mm]

$\beta$ Aur  &  A  & 3.930 $\pm$ 0.010 & A1V   & 2430. $\pm$ 250. & 9131 $\pm$ 257 \\
             &  B  & 3.962 $\pm$ 0.010 & A1V   & 2070. $\pm$ 139. & 9015 $\pm$ 182 \\[+1mm]

WW Aur       &  A  & 4.187 $\pm$ 0.019 & A5m   & 58.7  $\pm$ 8.65 & 7993 $\pm$ 470 \\
             &  B  & 4.143 $\pm$ 0.018 & A7m   & 49.3  $\pm$ 5.01 & 7651 $\pm$ 401 \\[+1mm]

PV Pup       &  A  & 4.257 $\pm$ 0.010 & A8V   & 22.0  $\pm$ 4.07 & 6870 $\pm$ 363 \\
             &  B  & 4.278 $\pm$ 0.011 & A8V   & 21.0  $\pm$ 2.41 & 6888 $\pm$ 265 \\[+1mm]

RS Cha       &  A  & 4.047 $\pm$ 0.023 & A8V   & 44.2  $\pm$ 5.56 & 7525 $\pm$ 307 \\
             &  B  & 3.961 $\pm$ 0.021 & A8V   & 43.8  $\pm$ 3.21 & 7178 $\pm$ 225 \\[+1mm]

HS Hya       &  A  & 4.326 $\pm$ 0.006 & F5V   & 7.55  $\pm$ 0.80 & 6398 $\pm$ 261 \\
             &  B  & 4.354 $\pm$ 0.006 & F5V   & 6.45  $\pm$ 0.40 & 6300 $\pm$ 217 \\[+1mm]

RZ Cha       &  A  & 3.909 $\pm$ 0.009 & F5V   & 7.50  $\pm$ 1.11 & 6440 $\pm$ 444 \\
             &  B  & 3.907 $\pm$ 0.010 & F5V   & 7.50  $\pm$ 0.60 & 6440 $\pm$ 396 \\[+1mm]

$\gamma$ Vir &  A  & 4.21  $\pm$ 0.017 & F0V   & 978.  $\pm$ 80.5 & 7143 $\pm$ 451 \\
             &  B  & 4.21  $\pm$ 0.017 & F0V   & 978.  $\pm$ 41.3 & 7143 $\pm$ 433 \\[+1mm]

DM Vir       &  A  & 4.108 $\pm$ 0.009 & F7V   & 3.00  $\pm$ 0.53 & 5931 $\pm$ 390 \\
             &  B  & 4.106 $\pm$ 0.009 & F7V   & 3.00  $\pm$ 0.28 & 5931 $\pm$ 319 \\[+1mm]

V624 Her     &  A  & 3.834 $\pm$ 0.010 & A3m   & 60.0  $\pm$ 6.44 & 8288 $\pm$ 476 \\
             &  B  & 4.024 $\pm$ 0.014 & A7V   & 29.0  $\pm$ 2.55 & 8092 $\pm$ 450 \\[+1mm]

V1143 Cyg    &  A  & 4.323 $\pm$ 0.016 & F5V   & 56.8  $\pm$ 6.32 & 6441 $\pm$ 201 \\
             &  B  & 4.324 $\pm$ 0.016 & F5V   & 53.2  $\pm$ 3.38 & 6393 $\pm$ 136 \\[+1mm]

MY Cyg       &  A  & 4.008 $\pm$ 0.021 & F0m   & 6.17  $\pm$ 0.76 & 7459 $\pm$ 891 \\
             &  B  & 4.014 $\pm$ 0.021 & F0m   & 6.00  $\pm$ 0.43 & 7408 $\pm$ 865 \\[+1mm]

$\delta$ Equ &  A  & 4.34  $\pm$ 0.02  & F7V   & 210.  $\pm$ 16.5 & 6393 $\pm$ 156 \\
             &  B  & 4.34  $\pm$ 0.02  & F7V   & 210.  $\pm$ 8.90 & 6393 $\pm$ 115 \\
\tableline\tableline
\end{tabular}
\end{table}

\section{Observations}

The H$\alpha$ and H$\beta$ observations were made at the Observatorio
del Roque de los Muchachos, La Palma using the Richardson-Brealey Spectrograph
on the 1.0m Jacobus Kapteyn Telescope (JKT) in October/November 1997. A 2400 l
${mm^{-1}}$ holographic grating was used and a 1124 $\times$ 1124 pixel Tek
CCD, giving a resolution of 0.4{\AA} {\sc fwhm}. Further H$\alpha$ and
H$\beta$ observations were made at the Mount Stromlo Observatory,
Australia in February 2000 using the Cassegrain Spectrograph on the ANU 74 inch
telescope. A 1200 l ${mm^{-1}}$ blazed grating was used,  giving a resolution
of 0.35{\AA} {\sc fwhm}. Additional H$\beta$ profiles were taken from the
work of Smalley \& Dworetsky (1995).

The data reduction of the profiles taken in 1997 and 2000 was performed using
the Starlink {\sc echomop} software package. In
most cases the final spectra had a signal-to-noise ratio in excess of 100:1.
Instrumental sensitivity variations were removed from the H$\alpha$ profiles by
comparing to observations of stars with intrinsically narrow Balmer profiles,
for example early-B or O type stars and G type stars, and the H$\beta$ profiles
were normalized such that the observed profile of Vega agreed to a model with
$T_{\mathrm eff}$=9550~K, $\log g$=3.95, [M/H]=$-$0.5 (Castelli \& Kurucz 1994) and the
standard profiles of Peterson (1969).

\section{Effective temperatures from Balmer line profiles}

The observed Balmer line profiles are fitted here to model spectra to compare
the derived $T_{\mathrm eff}$ with that from fundamental methods. The following
convection models were used, using solar-metallicity Kurucz {\sc atlas}
models:

\begin{description}

\item[MLT\_noOV 1.25] Standard {\sc atlas9} (Kurucz 1993) models using mixing length
theory (MLT) without convective overshooting. The value of the MLT parameter
$\alpha$ is the standard value of 1.25.

\item[MLT\_noOV 0.5] Standard {\sc atlas9} models using MLT without convective
overshooting. The value of the MLT parameter $\alpha$ is 0.5.

\item[MLT\_OV 1.25] Standard {\sc atlas9} models using MLT with approximate
overshooting. The value of the MLT parameter $\alpha$ used is 1.25.

\item[MLT\_OV 0.5] Standard {\sc atlas9} models using MLT with approximate convective
overshooting. The value of the MLT parameter $\alpha$ used is 0.5.

\item[CM] Modified {\sc atlas9} models using the Canuto \& Mazzitelli
(1991,1992) model of turbulent convection.

\end{description}

The synthetic spectra were calculated using {\sc uclsyn} (Smalley et al.
2001) which includes Balmer line profiles calculated using VCS Stark
broadening and metal absorption lines from the Kurucz \& Bell (1995)
linelist. This routine is based on the {\sc balmer} routine (Peterson 1969).
The synthetic spectra were normalized ${\pm100}$ \AA\  to match the
observations. The values of  $T_{\mathrm eff}$ were obtained by fitting model profiles
to the observations using the least-square differences.

\begin{figure}
\plotone{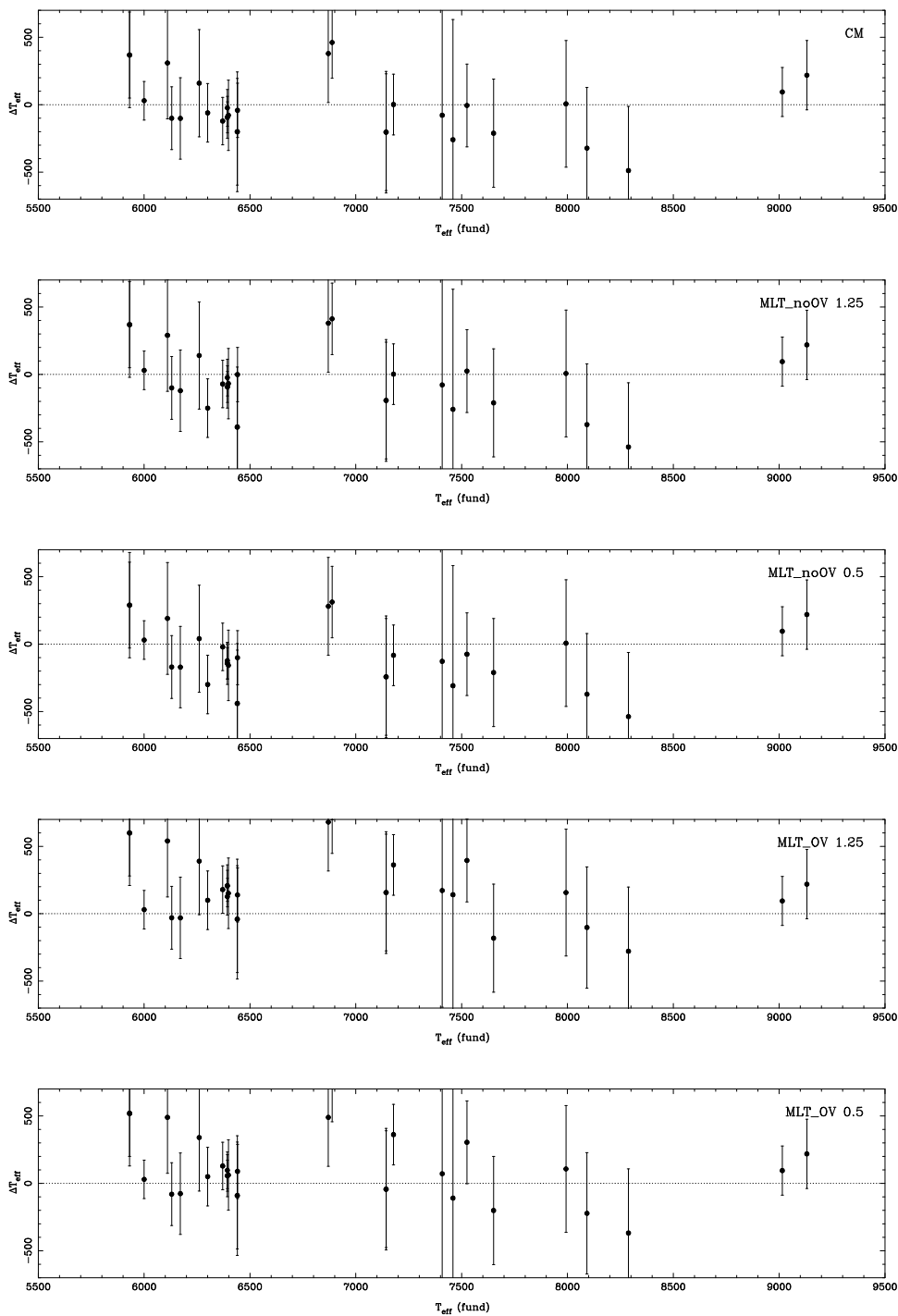}
\caption{Comparison between $T_{\mathrm eff}$ calculated from Balmer line
profiles H$\alpha$ to the Fundamental values.
$\Delta$$T_{\mathrm eff}$ $=$ $T_{\mathrm eff}$(Balmer) - $T_{\mathrm eff}$(fund) is
plotted against $T_{\mathrm eff}$(fund).}
\label{halpha_comp}
\end{figure}

\begin{figure}
\plotone{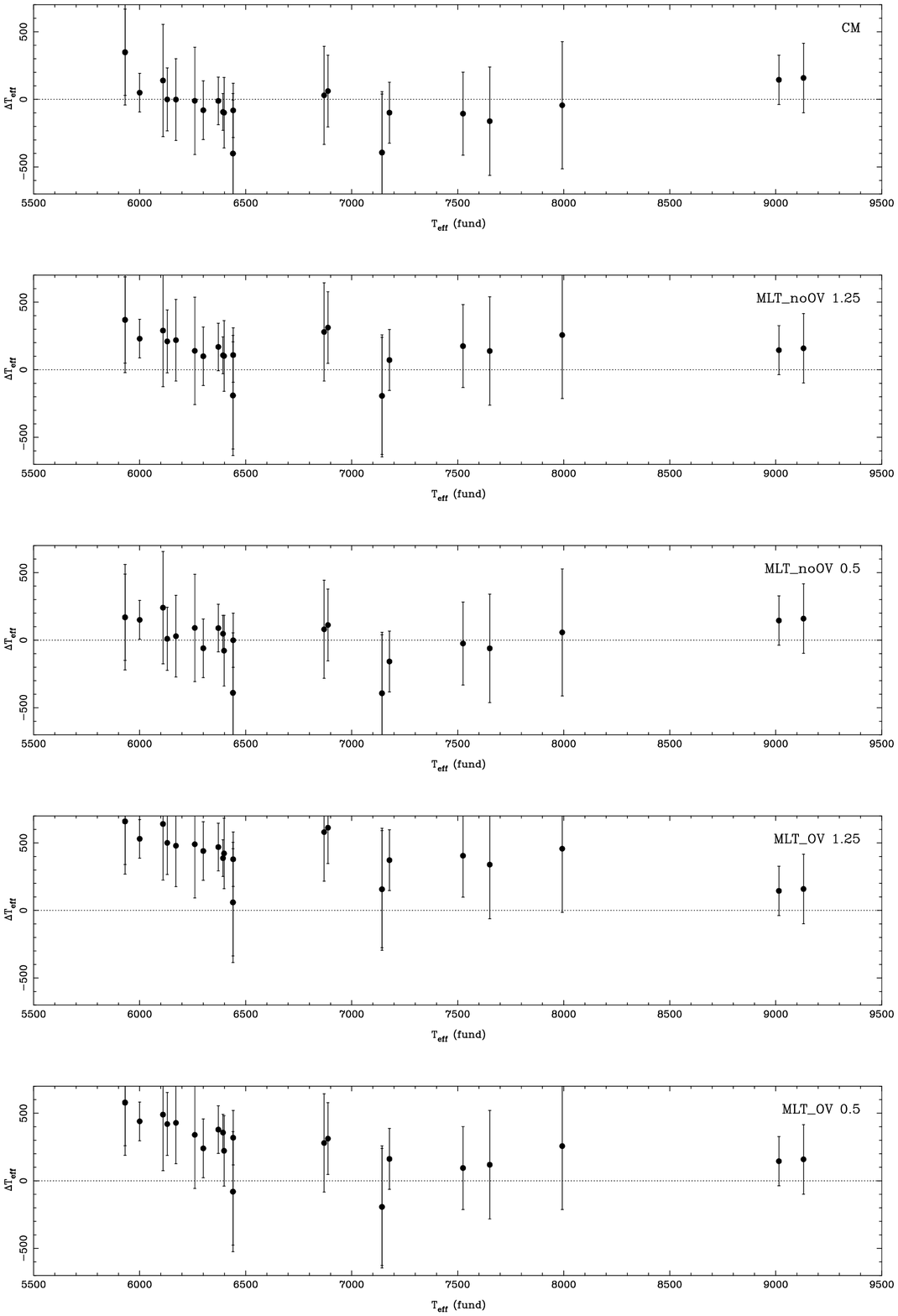}
\caption{Comparison between $T_{\mathrm eff}$ calculated from Balmer line
profiles H$\beta$ to the Fundamental values.
$\Delta$$T_{\mathrm eff}$ $=$ $T_{\mathrm eff}$(Balmer) - $T_{\mathrm eff}$(fund) is
plotted against $T_{\mathrm eff}$(fund). }
\label{hbeta_comp}
\end{figure}

Figures~1 \& 2 show the variation of $\Delta$$T_{\mathrm eff}$ = $T_{\mathrm eff}$(Balmer) $-$
$T_{\mathrm eff}$(fund) against $T_{\mathrm eff}$(fund) for H$\alpha$ and H$\beta$, respectively, for
the 5 convection models listed above. To within the uncertainties, the CM
results show no significant variation with $T_{\mathrm eff}$(fund) for either H$\alpha$ or
H$\beta$. The discrepancy around 8000~K noted by Gardiner et al. (1999) is not
evident. Even the two anomalous H$\alpha$ points just hotter than 8000~K, for
V624 Her, can be brought into agreement if the IRFM $T_{\mathrm eff}$ is used (Smalley
et al. 2002). The MLT\_noOV results are in broad agreement with those for CM,
but with the $\alpha$=0.5 models giving better agreement around 8000~K
relative to $\alpha$=1.25 and CM models.  Contrary to Gardiner et al. (1999),
who reported that F-type stars might require models with $\alpha \ge$1.25
(see their Fig.~9), we find that the binary systems do not support this.
Overall, $\alpha$=0.5 models are preferred to those with higher values. The
MLT\_OV models are generally more discrepant, yielding too high values of
$T_{\mathrm eff}$ (and even larger ones for H$\beta$, if $\alpha$=1.25 is used rather than
0.5), as found previously by Gardiner et al. (1999). Note also the systematic
difference between H$\alpha$ and H$\beta$ for $\alpha$=1.25 MLT\_noOV models,
which is even more pronounced for the MLT\_OV models.

\begin{table}
\caption{A-stars with fundamental values of $T_{\mathrm eff}$, but not $\log g$.}
\scriptsize    
\begin{tabular*}{\textwidth}{{@{\extracolsep{\fill}}l}*{10}{@{\extracolsep{\fill}}c}} \tableline
Star         & $v \sin i$    & $T_{\mathrm eff}$        & \multicolumn{2}{c}{CM $uvby$} & $T_{\mathrm eff}$ & \multicolumn{2}{c}{H$\alpha$} & \multicolumn{2}{c}{H$\beta$} \\
             & km s$^{-1}$   & Fund         & $T_{\mathrm eff}$    & $\log g$         & IRFM        & $T_{\mathrm eff}$ & $\log g$ & $T_{\mathrm eff}$ & $\log g$ \\ \tableline
$\gamma$ Gem &  45           & 9220$\pm$330 & 9250 & 3.56 & 9040$\pm$86 & 9220$\pm$300 & 3.40$\pm$0.2 & 9060$\pm$250 & 3.52$\pm$0.2 \\
$\beta$ Leo  & 122           & 8870$\pm$350 & 8770 & 4.32 & 8660$\pm$60 & 8370$\pm$400 & 3.77$\pm$0.2 & 8450$\pm$350 & 4.07$\pm$0.2 \\
$\alpha$ Oph & 240           & 7960$\pm$330 & 7940 & 3.80 & 7883$\pm$63 & 7510$\pm$100 & 3.69$\pm$0.3 & 7580$\pm$150 & 3.42$\pm$0.6 \\
$\alpha$ Aql & 245           & 7990$\pm$210 & 7840 & 4.18 & 7588$\pm$73 & 7420$\pm$100 & 4.17$\pm$0.3 & 7450$\pm$150 & 4.38$\pm$0.6 \\
$\alpha$ PsA &  85           & 8760$\pm$310 & 8890 & 4.30 & 8622$\pm$86 & 8340$\pm$400 & 3.87$\pm$0.2 &      &      \\
\tableline
\end{tabular*}
\label{Astars}
\end{table}

\section{The apparent A-star anomaly}

The use of stars with fundamental values of both $T_{\mathrm eff}$ and $\log g$ has failed
to support the apparent anomaly around 8000~K found by Gardiner et al.
(1999). However, there were too few stars within the $T_{\mathrm eff}$ range
8000--9000~K to fully explore this region. Gardiner et al. (1999) found that
four fundamental $T_{\mathrm eff}$ stars also showed the anomaly: $\beta$ Leo, $\alpha$
Oph, $\alpha$ Aql, $\alpha$ PsA. In order to be sure that there is no anomaly
in the Balmer line profiles, we need to explain why these stars might appear
anomalous.

Table~2 summarizes the values of $T_{\mathrm eff}$ obtained from CM $uvby$
photometry, the IRFM and by fitting to H$\alpha$ and H$\beta$ profiles. We have
allowed both $T_{\mathrm eff}$ and $\log g$ to vary in order to obtained the
best least-squares fit (see Figures~3 \& 4). Values of $\log g$ are also given
as obtained from $uvby$ photometry. We have also included $\gamma$ Gem which is
just hotter than 9000~K, but the results are in agreement with its
fundamental and IRFM $T_{\mathrm eff}$ values.

\begin{figure}
\plotone{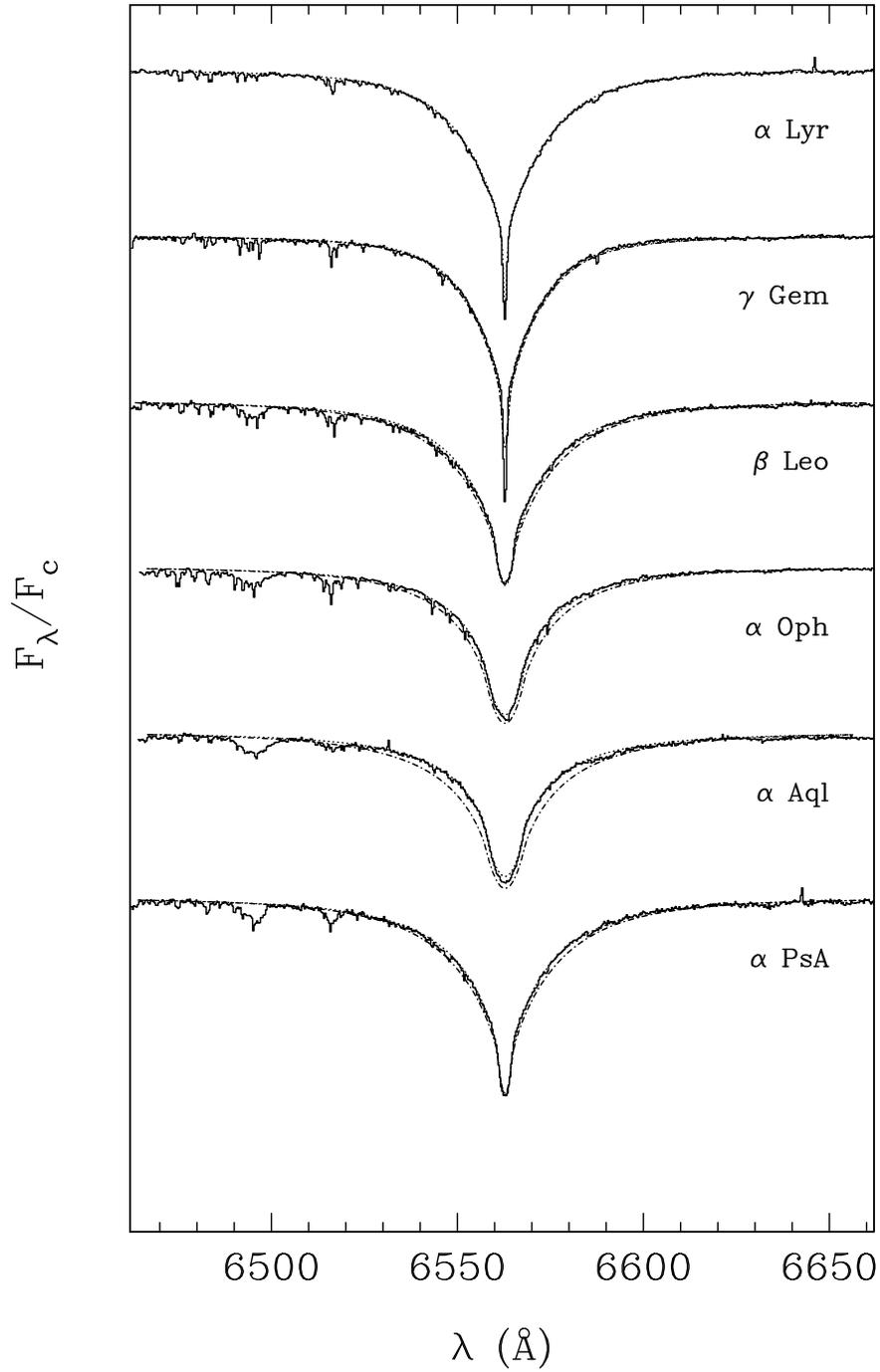}
\caption{H$\alpha$ profiles of A stars. The continuous line is the
observed profile, while the dotted line is the synthetic profile for
best fitting parameters given in Table~\ref{Astars}. The dash-dot
line is that for profiles calculated for the fundamental parameters.}
\label{astars-halpha}
\end{figure}

\begin{figure}
\plotone{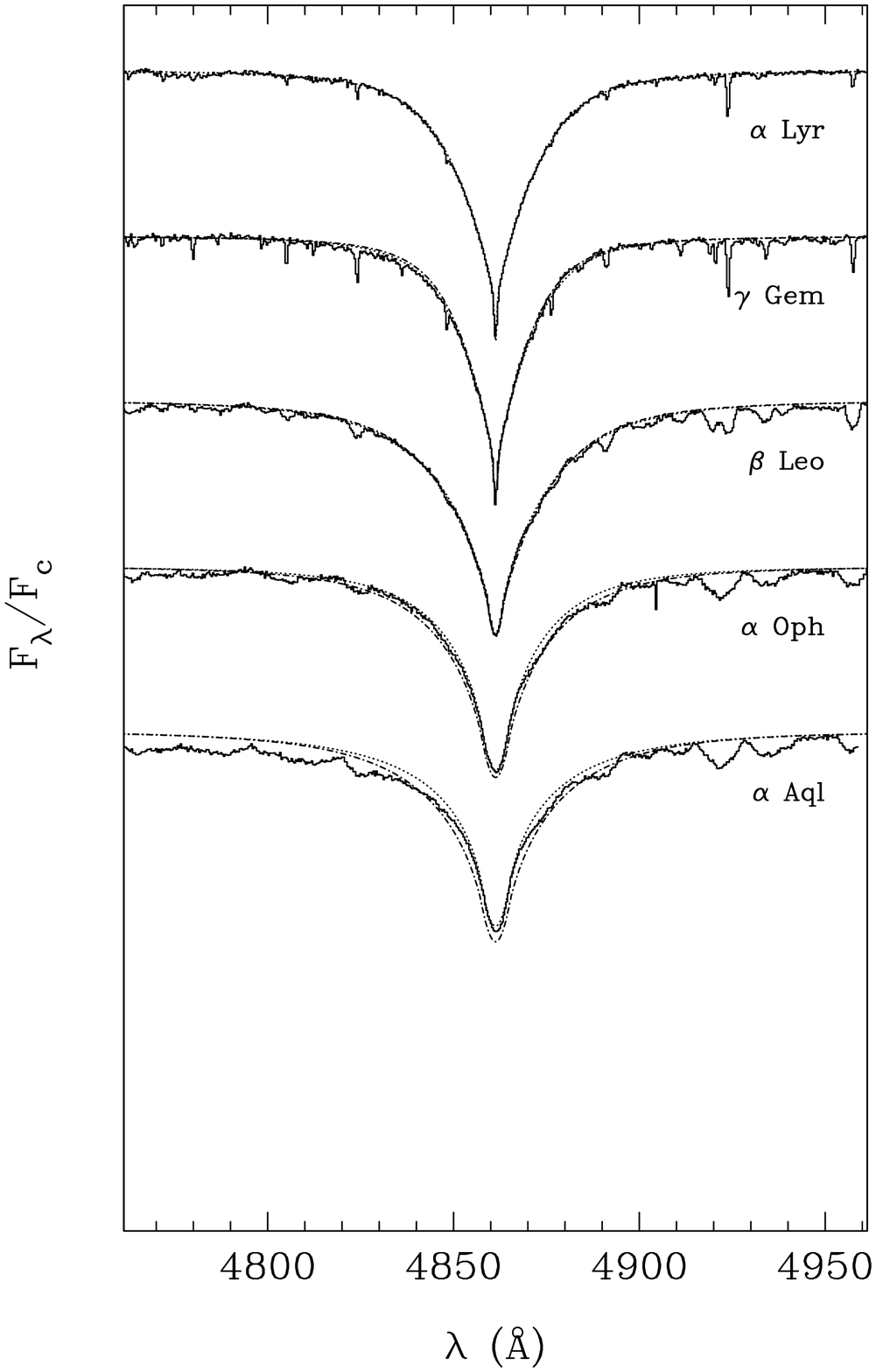}
\caption{H$\beta$ profiles of A stars. The continuous line is the
observed profile, while the dotted line is the synthetic profile for
best fitting parameters given in Table~\ref{Astars}. The dash-dot
line is that for profiles calculated for the fundamental parameters.}
\label{astars-hbeta}
\end{figure}

The rapidly rotating star $\alpha$ Aql has recently been studied by van Belle
et al. (2001) using interferometry. Their analysis revealed the oblateness
of the star and a new determination of fundamental $T_{\mathrm eff}$ = 7680$\pm$90~K.
This is significantly cooler than the previous determination, but in accord
with that inferred from the IRFM. As such, the $T_{\mathrm eff}$ from H$\alpha$ and
H$\beta$ are no longer significantly discrepant. It is certainly possible that
revision to the other fundamental stars could occur once new interferometric
measurements are obtained, especially $\alpha$ Oph which has a similar
$v \sin i$ and might be expected to exhibit significant oblateness. Thus, the
anomalies for these two stars can be explained in terms of their {\em rapid
rotation}.

The two other stars, $\beta$ Leo and $\alpha$ PsA, have lower $v \sin i$ values, but
are two most discrepant stars in the Gardiner et al. (1999) sample. Unless
the fundamental values are truly wrong there must be some other reason for
the discrepancy. The IRFM values both point to a slightly cooler $T_{\mathrm eff}$, but
even then the discrepancy is $\sim$300~K. However, in this temperature region
the Balmer lines are near their maximum strength and sensitive to $\log g$. It
is certainly possible that a small error in adopted $\log g$ could lead to a
large error in $T_{\mathrm eff}$ obtained from Balmer profiles. In addition, the Balmer
profiles change little with relatively large changes in $T_{\mathrm eff}$. Thus, we
conclude that the two stars are not discrepant, due to the low sensitivity of
Balmer lines with respect to changes in $T_{\mathrm eff}$ and both sensitivity to, and
the uncertainty in, the surface gravity for these stars. However, it must be
noted that the $\log g$ obtained from the Balmer lines for these stars is
systematically lower than that obtained from $uvby$  photometry.

In general, for stars hotter than 8000~K the sensitivity to $\log g$ prevents
us from using them to obtain values of $T_{\mathrm eff}$ to the accuracy required for
the present task, unless we have accurate fundamental values of $\log g$.
However, until we do have stars with accurate fundamental $\log g$ values, we
cannot be totally sure that there is not a problem with the model predictions
in this $T_{\mathrm eff}$ region.

\section{Conclusion}

Balmer line profiles have been fitted to the fundamental binary systems. To
within the errors of the fundamental $T_{\mathrm eff}$ values, neither the H$\alpha$ or
H$\beta$ profiles exhibit any significant discrepancies for the CM and MLT
without approximate overshooting models. As in previous work, the MLT with
overshooting models are found to be discrepant. Moreover, there are no
systematic trends, such as offsets, between results from H$\alpha$ and H$\beta$
as long as $\alpha$ in MLT models is chosen small enough (e.g. 0.5). The
discrepancies exhibited by the fundamental $T_{\mathrm eff}$ stars in Gardiner et al.
(1999) can be explained by rapid rotation in two cases and by the fact that
the Balmer profiles become sensitive to $\log g$ and less sensitive to $T_{\mathrm eff}$
in the other two cases. However, for the time being the lack of any stars
with fundamental values of both $T_{\mathrm eff}$ and $\log g$ in this region precludes
the conclusion that there is not a problem with the models in the $T_{\mathrm eff}$
range 8000 $\sim$ 9000~K.

Full details of this work are given in Smalley et al. (2002).

\acknowledgments
 
This work has made use of the hardware and software provided at Keele by the
PPARC Starlink Project. Friedrich Kupka acknowledges support by the project
{\em Turbulent convection models for stars}, grant P13936-TEC of the Austrian
Fonds zur F\"orderung der wissen\-schaft\-lichen Forschung.

\end{document}